\documentclass[%
 reprint,
 amsmath,amssymb,
 prx,
]{revtex4-2}

\usepackage{tikz}
\usetikzlibrary{quantikz}
\usetikzlibrary{shadings}
\usepackage{braket}
\usepackage{amsmath, amsthm, amssymb}%to use \split environment
\usepackage{bbm}
\theoremstyle{definition}%to define definition
\newtheorem{definition}{Definition}[section]

\newtheorem{theorem}{Theorem}[section]%theorem
\newtheorem{corollary}{Corollary}[section]%corollary
\newtheorem*{notation}{Notation}%Notation
\definecolor{bg1color}{RGB}{35, 63, 178}
\definecolor{bg2color}{RGB}{20, 140, 180}
\definecolor{bg3color}{RGB}{178, 100, 63}
\definecolor{bg4color}{RGB}{178, 150, 35}
\definecolor{bg5color}{RGB}{178, 100, 63}
\definecolor{darkgreen}{RGB}{28, 129, 0}
\begin{document}

\title{Near-optimal quantum circuit construction via Cartan decomposition}
\author{Maximilian Balthasar Mansky}%
 \email{maximilian-balthasar.mansky@ifi.lmu.de}
 \affiliation{LMU Munich}
 \author{Santiago Londoño Castillo}
 \affiliation{LMU Munich}
 \author{Victor Ramos Puigvert}
 \affiliation{LMU Munich}
\author{Claudia Linnhoff-Popien}
\affiliation{LMU Munich}
%\affiliation{%
%Lehrstuhl für mobile und verteilte Systeme, LMU Munich, Germany}

\date{\today}
\begin{abstract}
We show the applicability of the Cartan decomposition of Lie algebras to Quantum Circuits. This approach can be used  to synthesize circuits that can efficiently implement any desired unitary operation. Our method finds explicit quantum circuit representations of the algebraic generators of the relevant Lie algebras allowing the direct implementation of a Cartan decomposition on a quantum computer. The construction is recursive and allows us to expand any circuit down to generators and rotation matrices on individual qubits, where through our recursive algorithm we find that the generators themselves can be expressed with CNOT and SWAP gates explicitly. Our approach is independent of the standard CNOT implementation and can be easily adapted to other cross-qubit circuit elements. In addition to its versatility, we also achieve near-optimal counts when working with CNOT gates, achieving an asymptotic CNOT cost of $\frac{21}{16} 4^n$  for $n$ qubits.
\end{abstract}

\maketitle

\section{Introduction}
\label{sec:introduction}

Quantum computing relies on a quantum circuit to translate an algorithm to work on a quantum computer. The circuit expresses the physical actions that are necessary to create a particular quantum-mechanical state and whose measurement provides the output of the calculation. 
%
%In the current NISQ era, the length of a circuit is an important success parameter for a quantum computation. Every operation on the qubits carries a chance of an error, therefore optimizing the number of operations on each qubit is an important success strategy. A number of methods for circuit optimization have been introduced recently \cite{barenco1995elementary, knill1995approximation, vartiainen2004efficient, mottonen2004quantum, shende2005synthesis}, with varying degrees of optimality.
%
Every circuit is equivalent to a unitary transformation in $SU(2^n)$, where $n$ refers to the number of qubits. The mapping is injective, in the sense that two different circuits can perform the same calculation and correspond to the same transformation $U$ \cite{nielsen_quantum_2002}. Consequently, circuits of different lengths can perform the same circuit. In most cases, the shorter circuit is preferable, since it reduces the execution time, imprecision due to hardware limitations or, on noisy systems, reductions of noise due to fewer actual operations. Various methods can be employed to optimize circuits \cite{gabor_applicability_2022, knill1995approximation, hietala_verified_2021}; however, before optimization, the circuit must first be constructed.

There are several ways to construct quantum circuits. One can construct an algorithm along a schema – the well-known algorithms of Shor \cite{shor_algorithms_1994} and Grover \cite{grover_fast_1996} work in this way and can be scaled to the required system size by following the schema. Finding new schematic algorithms and showing their speed-up compared with classical methods is its own field of research. So far, the number of discovered algorithms is limited \cite{shor_progress_2004, shao_quantum_2019}.

An alternative is to use a parametrized circuit and modify the parameters until the circuit fits the desired output. The approach is called quantum machine learning  \cite{schuld_introduction_2015, biamonte_quantum_2017, cerezo_challenges_2022}. Similar to classical machine learning, some quantum circuit ansatz is chosen, often with distinct layers of repeated subcircuits, which is then trained with some classical feedback loop to approximate a desired solution.

In our work we provide a solution to a third approach, decomposing a known unitary matrix into its corresponding quantum circuit. We can build circuits for any arbitrary target, not just the ones for which we have schemas, and also with known performance, as we know the number of required CNOT gates. Our approach provides a direct method for translating a unitary operation $U$ to an explicit quantum circuit.

The developed algorithm generalizes the unstructured circuit decomposition of a three-qubit unitary, done in \cite{vatan2004realization}, to an $n$-qubit unitary by using a recursive method. The mathematics upon which this recursive algorithm is constructed is based  on the work of Khaneja and Glaser \cite{khaneja_cartan_2001}. Underlying our construction is the Cartan decomposition of a unitary $U\in SU(2^n)$ 
into four terms $K_1 \exp(z_1) K_2 \exp(y) K_3 \exp(z_2) K_4$, where all $K_i$ are part of the next-lower dimension group $K_i \in SU(2^{n-1}) \otimes U(1)$, and $z_i$ and $y$ are algebra elements belonging to certain Cartan subalgebras. We show that this argument is recursive and allows us to decompose any unitary into components that can be easily represented in a quantum circuit. This is described in section \ref{sec:khaneja-glaser-decomposition}. 

The orthogonal elements $\exp(z_i)$ and $\exp(y)$ in the Cartan decomposition are created through the generators of the Lie subalgebras and will ultimately contain the only cross-qubit elements in the circuit. To express them in terms of circuit elements, we make use of a block diagonal decomposition to the elements of the algebra. This form is easily expressible in terms of CNOTs and elementary rotations, described in detail in section \ref{sec:block-diagonal-decomposition}. 

We also assess the performance of our algorithm as expressed by the number of CNOTs that an arbitrary circuit requires in the worst case. The number of gates can be determined analytically, see section \ref{sec:cnot_gates}. There we also compare our CNOT count to other methods decomposing a unitary \cite{barenco1995elementary, knill1995approximation, vartiainen2004efficient, mottonen2004quantum, shende2005synthesis} and we also provide an outlook of future work in section \ref{sec:Discussion}.

%The remainder of this paper is structured as follows. In \ref{sec:khaneja-glaser-decomposition} we introduce the Khaneja-Glaser decomposition and all the necessary mathematical background. In \ref{sec:block-diagonal-decomposition} we present the main decomposition method of this paper, where all the generators of the Cartan subalgebras get decomposed through quantum circuits converting them to block diagonal matrices. In \ref{sec:example} we give an explicit example of how a unitary $U\in SU(8)$ gets decomposed through this Cartan block-diagonal decomposition. In \ref{sec:cnot_gates} we determine the cost of CNOT gates required to implement the decomposition of a unitary in $SU(2^n)$. In \ref{sec:comparison_with_other_methods} we compare how the Khaneja-Glaser decomposition performs with respect to other algorithms. In \ref{sec:future_work} we outline some of the possible future work

\begin{figure*}[hbt]
\includegraphics{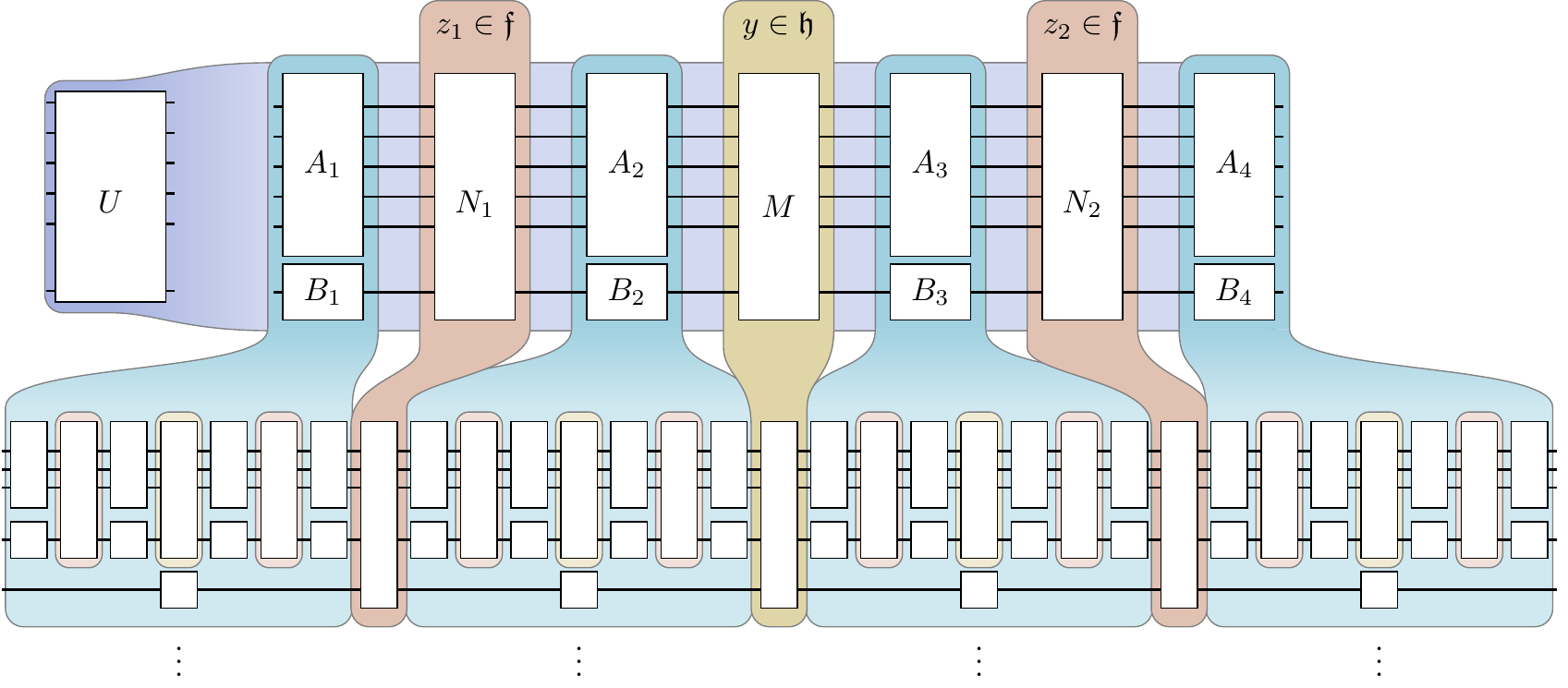}
\caption{
Graphical representation of our work. We establish a structure from an arbitrary unitary $U$ into circuit elements of lower dimension ($K_i = A_i \otimes B_i$, $A_i\in SU(2^{n-1})$, $B_i \in U(1)$) and $n$-qubit elements that generate from the algebras $\mathfrak{f}$ and $\mathfrak{h}$. The algorithm is recursive for all $A_i$ and detailed in section \ref{sec:khaneja-glaser-decomposition}. Between the recursive elements are the $(n-k)$-qubit elements, where $k$ refers to the number of recursions. The elements can be expressed explicitly as CNOT and SWAP elements through the block diagonal decomposition explained in section \ref{sec:block-diagonal-decomposition}.}
%Graphical representation of the main results of our work. We can decompose an arbitrary operation $U \in SU(2^n)$ (blue background) into group elements $K_i = A_i \otimes B_i \in SU(2^{n-1}) \otimes U(1)$ (teal background) and elements of the algebra $N_i$, $M$ (i), (brown and gold backgrounds). We detail this approach in section \ref{sec:khaneja-glaser-decomposition}. The approach is recursive, and we can repeatedly decompose the elements $A_i \in SU(2^{n-1})$ into smaller and smaller subspaces. The last step on $SU(4)$ is slightly different and decomposes into one algebra element and rotation matrices (ii). We can then express each many-qubit element in terms of CNOTs and SWAPs with a finite number of generators (iii), as detailed in section \ref{sec:block-diagonal-decomposition}.}\label{fig:decomposition-representation}
\label{fig:decomposition-representation}
\end{figure*}

\section{Related work}
We provide a general overview of some of the most relevant algorithms for the synthesis of general multi-qubit gates; Cosine-Sine decomposition \cite{mottonen2004quantum}, Optimized Quantum Shannon Decomposition (QSD) \cite{shende2005synthesis}, and Khaneja-Glaser decomposition \cite{khaneja_cartan_2001}. The theory behind the Khaneja-Glaser decomposition is discussed in more detail, as our work relies on the mathematical structure and extends their work to an arbitrary number of qubits.

\subsection{Cosine-sine decompositon}
\label{sec:CSDsection}
One way to realize a general $\text{SU}(2^n)$ matrix on a quantum computer is via matrix factorization, where the initial matrix is separated into a product of matrices which can be more easily implemented as a quantum circuit. Such a factorization can be recursively achieved by using the Cosine-Sine Decomposition (CSD) \cite{PAIGE1994303}. In general, the CSD of a SU$(2^n)$ matrix $U$ can be written as follows:

\begin{equation}\label{eq:CSD}
\resizebox{0.9 \columnwidth}{!}{$
U=U_1^1 A_1^1 \Tilde{U}^1_2=
\begin{pmatrix}
u^1_{11} & 0 \\
0 & u^1_{12}  
\end{pmatrix}
\begin{pmatrix}
c^1_{11} & s^1_{11} \\
-s^1_{11} & c^1_{11}  
\end{pmatrix}
\begin{pmatrix}
\Tilde{u}^1_{21} & 0 \\
0 & \Tilde{u}^1_{22}  
\end{pmatrix}
$%
}
\end{equation}

This decomposition can be applied recursively to the sub-matrices $U^i_j$ until a $2 \times 2$ block-diagonal form is obtained.

In \cite{PAIGE1994303}, it is shown that the matrices resulting from the above decomposition can be attained as a product of uniformly controlled rotations. After canceling some of the occurring $\operatorname{CNOT}$ gates using reflection symmetries of the circuit, and using a method for implementing uniformly controlled gates described in the paper, the authors show that a general CSD of a SU$(2^n)$ matrix, as shown in equation \eqref{eq:CSD}, can be implemented using $4^n-2^{n+1}$ $\operatorname{CNOT}$ gates and $4^n$ one-qubit gates.

\subsection{Optimized quantum Shannon decomposition}

Another way to decompose a generic unitary matrix is by generalizing the concepts of Boolean algebra and logic conditionals to quantum circuits. By interpreting the qubits as the predicates and requiring the action of clauses to be unitary, operations in a quantum circuit can then be interpreted as quantum conditionals. In \cite{shende2005synthesis}, the authors introduce \textit{quantum multiplexors} as quantum circuit blocks implementing quantum conditionals, e.g. the $\operatorname{CNOT}$ gate is the simplest 2-qubit multiplexor. To perform the decomposition of a unitary matrix, the authors provide a generalization to quantum circuits of the classical Shannon Decomposition  theorem, which allows any Boolean function $F$ to be factorized as $F= x \cdot F_x + \Bar{x} \cdot F_{\Bar{x}}$, where $x$ is a variable and $\Bar{x}$ its complement. The proposed Quantum Shannon Decomposition (QSD) theorem states that an arbitrary $n$-qubit operator can be implemented by a circuit containing three multiplexed rotations and four generic $(n-1)$-qubit operators. This provides a method to recursively decompose a generic SU$(2^n)$ operator. Applying this theorem to the previously discussed CSD, see \ref{sec:CSDsection}, and by providing a method to implement multiplexed-$R_y$ rotations using $\operatorname{Controlled-Z}$ gates, the authors showed that the number of $\operatorname{CNOT}$ gates required to decompose a  SU$(2^n)$ matrix can be reduced to $\frac{23}{48} 4^n-\frac{3}{2}2^n+ \frac{4}{3}$, a significant improvement from the previously discussed CSD.

Both approaches use post-circuit creation optimization to improve their count of operations. We compare the achieved CNOT counts with our own in table \ref{tab:comparison_CNOT}.

\section{Cartan Decomposition}

The Cartan decomposition method is a powerful tool in the realm of Lie group decomposition. It allows us to break down a given Lie group into smaller, simpler subgroups, which can be much easier to work with. This method has found numerous applications in a variety of fields, including physics, engineering, and computer science. Building upon the work of Khaneja and Glaser, we have extended their method, which uses the Cartan decomposition of a Lie group, to be applicable to an arbitrary system size. 

\subsection{Khaneja-Glaser decomposition}
\label{sec:khaneja-glaser-decomposition}

The underlying mathematics here relies on the work of Elie Cartan \cite{cartan_sur_1926, cartan_sur_1927} in French and is by now part of the standard knowledge of physics and mathematics. For an English language introduction to Lie groups and algebras, see e.g. \cite{humphreys_introduction_2012}. Throughout this paper, let $G$ be a compact semi-simple Lie group with identity $e$ and let $\mathfrak{g}$ denote its Lie algebra. Moreover, let $K$ denote a compact closed subgroup of $G$. Note that, given that $\mathfrak{g}$ is a semi-simple algebra there exists, due to Cartan's criterion, a non-degenerate Killing form inducing a bi-invariant metric $\langle\cdot,\cdot\rangle_G$ on $G$, which allows the sum decomposition of $\mathfrak{g}$ into subalgebras. 

\begin{notation}
Throughout this paper, capital letters identify groups. Capital letters with subscripts identify elements of the group. The algebras are denoted by lowercase fraktur letters, elements thereof by lowercase letters. Pauli matrices are referenced by their standard $\sigma_i$. We also make use of the following notation for generalized Pauli matrices 
\begin{equation*}	x_k \equiv \mathbbm{1}\otimes\cdots\otimes\mathbbm{1}\otimes\sigma_x\otimes\mathbbm{1}\otimes\cdots\otimes\mathbbm{1}
\end{equation*}
where the Pauli matrix $\sigma_x$ acts on the $k$th qubit. Matrices for the rotations around $\sigma_y$ and $\sigma_z$ are constructed similarly and denoted $y_k$ and $z_k$ respectively.
\end{notation}

\begin{definition}[Cartan decomposition of $\mathfrak{g}$]
Let $\mathfrak{g}$ and $\mathfrak{l}$ be the two real semi-simple Lie algebras of $G$ and $K$ respectively. Then, $(\mathfrak{g},\mathfrak{l})$ is called an \textbf{orthogonal symmetric Lie algebra pair} if the decomposition $\mathfrak{g}=\mathfrak{m}\oplus\mathfrak{l}$, where $\mathfrak{m}=\mathfrak{l}^\perp$, satisfies the following commutation relations
\begin{itemize}
    \item[(i)] $[\mathfrak{l},\mathfrak{l}]\subset\mathfrak{l}$,
    \item[(ii)] $[\mathfrak{m},\mathfrak{l}]=\mathfrak{m}$, 
    \item[(iii)] $[\mathfrak{m},\mathfrak{m}]\subset\mathfrak{l}$.
\end{itemize}
The direct sum decomposition $\mathfrak{g}=\mathfrak{m}\oplus\mathfrak{l}$ is then called a Cartan decomposition of the Lie algebra $\mathfrak{g}$.
\end{definition}

\begin{definition}[Cartan subalgebra]\label{def:cartan-subalgebra}
Let $(\mathfrak{g},\mathfrak{l})$ be an orthogonal symmetric Lie algebra pair of the groups $G$ and $K$. A maximal subalgebra $\mathfrak{h}$ of $\mathfrak{m}$ is called a \textbf{Cartan subalgebra} of $(\mathfrak{g},\mathfrak{l})$.
\end{definition}
In \cite{khaneja_cartan_2001} it is shown that the Lie algebra $\mathfrak{su}(2^n)$ defined by
\begin{align*}
\mathfrak{su}(2^n)=&\text{span}\{a\otimes\sigma_x,b\otimes\sigma_y,c\otimes\sigma_z,d\otimes \mathbbm{1},\\
&ix_n,iy_n,iz_n\vert a,b,c,d\in \mathfrak{su}(2^{n-1})\}
\end{align*}
has a Cartan decomposition $\mathfrak{su}(2^n)=\mathfrak{su}_{\mathfrak{m}}(2^n)\oplus\mathfrak{su}_{\mathfrak{l}}(2^n)$, where
\begin{align*}
    \mathfrak{su}_{\mathfrak{m}}(2^n)&=\text{span}\{a\otimes\sigma_x,b\otimes\sigma_y,ix_n,iy_n\vert a,b\in \mathfrak{su}(2^{n-1})\}\\
    \mathfrak{su}_{\mathfrak{l}}(2^n)&=\text{span}\{c\otimes\sigma_z,d\otimes\mathbbm{1},iz_n\vert c,d\in\mathfrak{su}(2^{n-1})\}.
\end{align*}

\begin{theorem}[Cartan decomposition of $G$]
Let $\mathfrak{g}$ be a semi-simple Lie algebra of the group $G$ and let $\mathfrak{g}=\mathfrak{m}\oplus\mathfrak{l}$ be its Cartan decomposition. Moreover, let $\mathfrak{h}$ be a Cartan subalgebra of $(\mathfrak{g},\mathfrak{l})$ and let $K$ be a compact closed subgroup of $G$. Then, 
\begin{equation}
    G=K\exp(\mathfrak{h})K,
    \label{eq:cartandecomposition}
\end{equation}
where $\exp{(\mathfrak{h})}\subset G$. This decomposition is then called the \textbf{Cartan decomposition} of the Lie group $G$.
\end{theorem}

The decomposition of the Lie group $G$ into two groups $K$ linked by a determinable element of the algebra is the heart of our algorithm. In terms of the actual elements of the group, we obtain a structure as below.

\begin{corollary}
Let $U\in SU(2^n)$ be an $n$-qubit unitary operator. Then it has a decomposition

\begin{equation}
    U=K_1\exp(y)K_2
    \label{eq:cartandecomposition_corollary}
\end{equation}
where $K_i\in  \exp(\mathfrak{su}_{\mathfrak{l}}(2^n))$ and for some $y \in\mathfrak{h}$, where $\mathfrak{h}$ is a Cartan subalgebra of $(\mathfrak{su}(2^n),\mathfrak{su}_{\mathfrak{l}}(2^n))$.
\end{corollary}

In \cite{khaneja_cartan_2001} it is proven that $K_i\in  \exp(\mathfrak{su}_{\mathfrak{l}}(2^n)) \cong SU(2^{n-1})\otimes SU(2^{n-1})\otimes U(1)$ so that the unitaries $K_i$ have again a Cartan decomposition. This provides a recursive algorithm for determining a unitary $U\in SU(2^n)$ by successive decompositions.
\begin{theorem}
The direct sum decomposition
\begin{equation}
 \mathfrak{su}_{\mathfrak{l}}(2^n)=\mathfrak{su}_{\mathfrak{l}0}(2^n)\oplus\mathfrak{su}_{\mathfrak{l}1}(2^n), 
\end{equation}
where
\begin{equation*}
\begin{split}
    \mathfrak{su}_{\mathfrak{l}0}(2^n)=\text{span}\{c\otimes\sigma_z\vert c\in \mathfrak{su}(2^{n-1})\}\\
    \mathfrak{su}_{\mathfrak{l}1}(2^n)=\text{span}\{d\otimes\mathbbm{1},iz_n\vert d\in\mathfrak{su}(2^{n-1})\}
\end{split}
\end{equation*}
is a Cartan decomposition of the Lie algebra $\mathfrak{su}_{\mathfrak{l}}(2^n)$.
\end{theorem}
The proof of this theorem can also be found in \cite{khaneja_cartan_2001}.
%This decomposition is maximal in the sense that it separates the algebra $\mathfrak{su}(2^n)$ into two separate spaces of dimension $2^{n-1}$. 

\begin{corollary}\label{cor:2group}
Let $V\in \exp(\mathfrak{su}_l(2^n))$ be an n-qubit operator. Then it has a unique decomposition

\begin{equation}
    V=K_1\exp(z)K_2
\end{equation}
where $K_i\in SU(2^{n-1}) \otimes U(1)$ and for some $z\in \mathfrak{f}$, where $\mathfrak{f}$ is a Cartan subalgebra of $(\mathfrak{su}_{\mathfrak{l}}(2^n),\mathfrak{su}_{\mathfrak{l}0}(2^n))$.
\end{corollary}

\begin{corollary}\label{cor:4group}
Let $U\in SU(2^n)$ be an $n$-qubit unitary operator. Then it has a decomposition
\begin{equation}
    U=K_1\exp(z_1)K_2\exp(y)K_3\exp(z_2)K_4
    \label{eq:cartandecomposition_corollary2}
\end{equation}
where $K_i\equiv A_i\otimes B_i\in SU(2^{n-1})\otimes U(1)$, $y\in\mathfrak{h}$ and $z_i\in\mathfrak{f}$, where $\mathfrak{h}$ is a Cartan subalgebra of $(\mathfrak{su}(2^n),\mathfrak{su}_l(2^n))$  and $\mathfrak{f}$ is a Cartan subalgebra of $(\mathfrak{su}_l(2^n),\mathfrak{su}_{l0}(2^n))$.
\end{corollary}

%\begin{figure}[htb]
%	\centering
%	\includegraphics[width=.5\textwidth]{SU(16).pdf}
%	\caption{Circuit decomposition of a $n$-qubit unitary operation $U\in SU(2^n)$ using Khaneja-Glaser's decomposition, where $K_i\equiv A_i\otimes B_i\in SU(2^{n-1})\otimes U(1)$, $M\equiv e^y, y\in\mathfrak{h}$, and $N_i\equiv e^{z_i}, z_i\in\mathfrak{f}$, where $\mathfrak{h}$ is a Cartan subalgebra of $(\mathfrak{su}(2^n),\mathfrak{su}_{\mathfrak{l}}(2^n))$ and $\mathfrak{f}$ is a Cartan subalgebra of $(\mathfrak{su}_{\mathfrak{l}}(2^n),\mathfrak{su}_{\mathfrak{l}0}(2^n))$.}
%	\label{fig:KG}
%\end{figure}
In order to define a Cartan subalgebra in the product operator basis for the pairs $(\mathfrak{su}(2^n),\mathfrak{su}_{l}(2^n))$ and $(\mathfrak{su}_l(2^n),\mathfrak{su}_{l0}(2^n))$ we proceed analogously as in \cite{khaneja_cartan_2001}. The elements of the Cartan subalgebra can be generated recursively by the following equations:
\begin{equation}
	\begin{split}
		\mathfrak{a}(2)&=i\{x_1x_2,y_1y_2,z_1z_2\}, \quad \mathfrak{b}(2)=\varnothing \\
		\mathfrak{s}(k)&=\bigcup_{i=2}^{k}\mathfrak{a}(i) \otimes \mathbbm{1}^{k-i} \\
		\mathfrak{a}(n)&=\{\alpha\otimes \sigma_x, ix_n\vert \alpha \in\mathfrak{s}(n-1)\}\\
		\mathfrak{b}(n)&=\{\alpha \otimes\sigma_z\vert \alpha \in\mathfrak{s}(n-1)\}\\
		\mathfrak{h}(n)&=\text{span}\{\hspace{0.4ex}\mathfrak{a}(n)\}\\
		\mathfrak{f}(n)&=\text{span}\{\hspace{0.4ex}\mathfrak{b}(n) \}
		\label{eq:recursion}
	\end{split}
\end{equation}
This decomposition structure allows us to express any $n$-qubit unitary in terms of ($n-1$)-qubit unitaries and elements of orthogonal algebras. The circuit structure is visualized in figure \ref{fig:decomposition-representation}. Recursively it follows that each of the $A_i$ shown in the figure can itself be decomposed in the same way. This decomposition method of an $n$-qubit unitary works all the way down to $SU(4)$, the space of two-qubit operations, which can be further decomposed by $K_1 \exp(y) K_2$, where $K_i\in SU(2)$ and $y\in\mathfrak{h}(2)$ since $\mathfrak{f}(2)=\varnothing$.

It is important to note that $x_1x_2$ and so on are elements of the algebra, not elements of the group such as $X_1 \otimes X_2$. The corresponding group element $\exp(x_1 x_2)$ is not the direct product of two $X$ rotations but rather a two-qubit operation. Moreover, note that, for $\alpha \in\mathfrak{s}(n-1)$, $\alpha \otimes\sigma_x$ and $\alpha \sigma_{nx}$ represent the same element, where $\sigma_{xn}\equiv x_n$.

\section{Block Diagonal Decomposition}\label{sec:block-diagonal-decomposition}

By employing a recursive method, the developed algorithm extends the unstructured circuit decomposition of a three-qubit unitary, as demonstrated in \cite{vatan2004realization}, to an $n$-qubit unitary. This algorithm determines a decomposition for the generators $y\in\mathfrak{h}(n)$ and $z\in\mathfrak{f}(n)$ of the relevant Lie subalgebras $(\mathfrak{su}(2^n),\mathfrak{su}_l(2^n))$ and $(\mathfrak{su}_l(2^n),\mathfrak{su}_{l0}(2^n))$ using a block-diagonal matrix.

It is important to note that within these Lie subalgebras, there are always two generators constructed through the recursive equations in \eqref{eq:recursion} that are proportional to each other, since $x_1x_2$ is proportional to $y_1y_2$, and $z_1z_2$ is proportional to the identity. Consequently, in order to decompose the Cartan subalgebras $M=e^y$ and $N=e^z$, where $y\in\mathfrak{h}(n)$ and $z\in\mathfrak{f}(n)$ represent, respectively, the generators of the Cartan subalgebras $(\mathfrak{su}(2^n),\mathfrak{su}_l(2^n))$ and $(\mathfrak{su}_l(2^n),\mathfrak{su}_{l0}(2^n))$, we group the proportional generators together and separate the exponential terms into $2^{n-2}$ different components $M_i(a_i, b_i)$ and $N_i(a_i, b_i)$:
\begin{align}
    M=e^y\equiv &M_1(a_1,b_1)\cdots M_{2^{n-2}}(a_{2^{n-2}},b_{2^{n-2}})\\
    N=e^z\equiv N_1({a}_1)&N_2({a}_2,{b}_2)\cdots N_{2^{n-2}}({a}_{2^{n-2}},{b}_{2^{n-2}})
\end{align}
%\begin{align}
%    M=e^y\equiv M_1(a_1,b_1)\cdots M_{2^{n-2}}(a_{2^{n-2}},b_{2^{n-2}})\\
%    N=e^z\equiv N_1(a_{2^{n-2}+1},b_{2^{n-2}+1})\cdots N_{2^{n-2}}(c_{2^{n-1}},d_{2^{n-1}})
%\end{align}
An efficient way of mapping the generators to circuit elements is by means of a particular block-diagonal form:

\begin{equation}\label{eq:block-form}
P_\mp(a_i,b_i)=\operatorname{diag}(p_\mp,...,p_\mp,p_\pm,...,p_\pm) 
\end{equation}
with entries
\begin{equation}
\begin{split}
p_-=\begin{pmatrix}
		\cos(a_i-b_i)&i\sin(a_i-b_i)\\i\sin(a_i-b_i)&\cos(a_i-b_i)
	\end{pmatrix}\\
p_+=\begin{pmatrix}
	\cos(a_i+b_i)&i\sin(a_i+b_i)\\i\sin(a_i+b_i)&\cos(a_i+b_i)
	\end{pmatrix}
\end{split}
\end{equation}

\begin{figure}[h]
	\centering
	\includegraphics{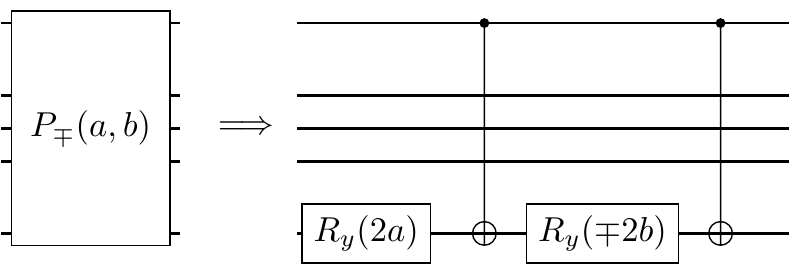}
	\caption{Quantum circuit decomposing the unitary block-form $P_\mp(a,b)$ up to phase. The block-diagonal matrices $P_\mp(a,b)$ can always be decomposed up to a global phase by a dimensionally adapted quantum circuit, where the one-qubit gates and the target qubits have to be adjusted accordingly.}
	\label{fig:block_form}
\end{figure}

The block-diagonal structure can be implemented on a quantum circuit in a straightforward way, visualized in figure \ref{fig:block_form}. The two parameters are implemented as rotation gates on the $n$th wire and controlled via CNOTs from the first.

\begin{figure*}[htb]
\includegraphics{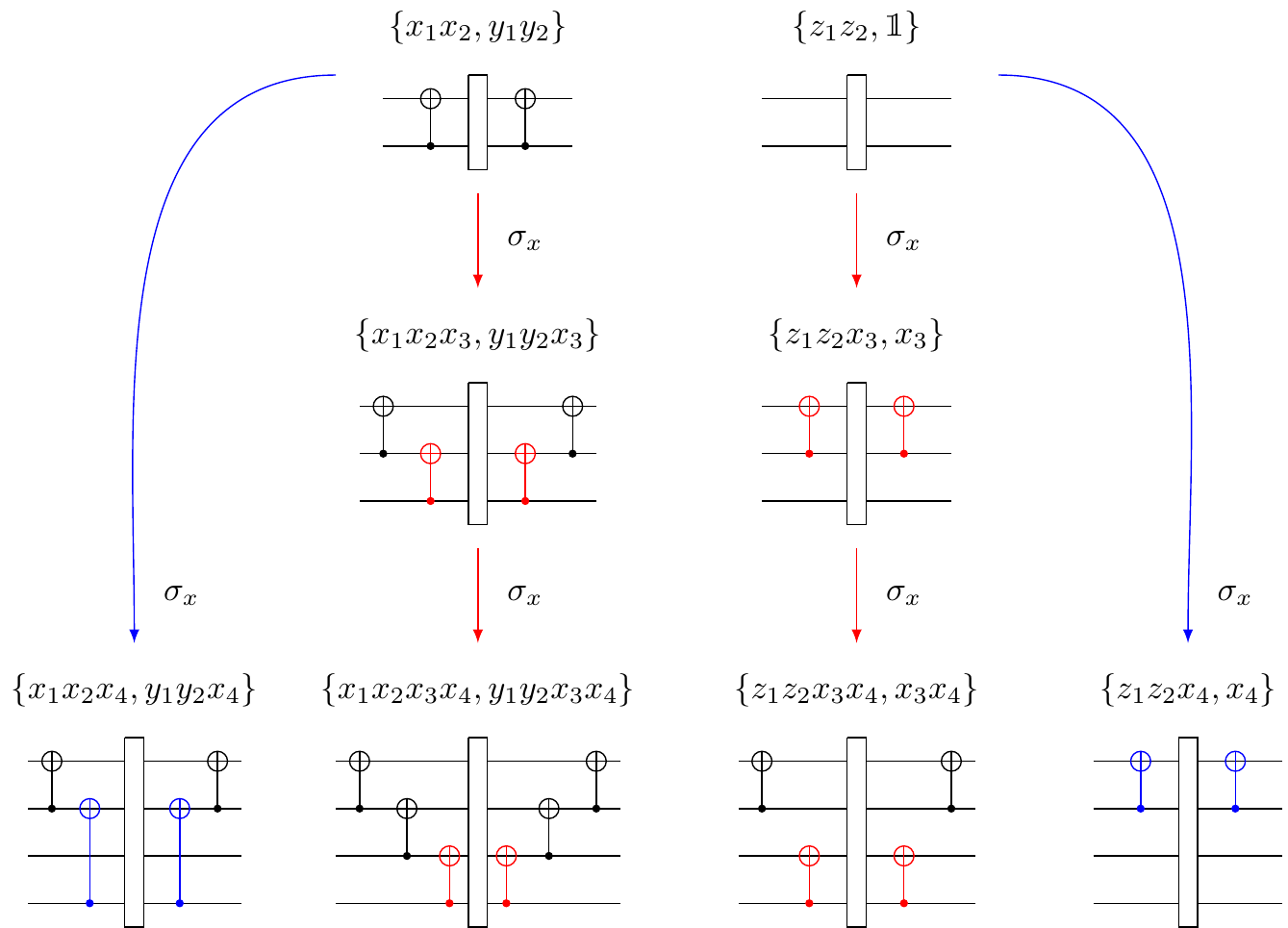}
	\caption{Recursive algorithm showing the decomposition of the exponential operator $M=e^y$, $y\in\mathfrak{h}(4)$, where $\mathfrak{h}(4)$ is a Cartan subalgebra of $(\mathfrak{su}(16),\mathfrak{su}_{\mathfrak{l}}(16))$. Each of the exponential operators $M_i(a_i,b_i)$ is decomposed through a circuit including a block-diagonal matrix $P_\mp(a_i,b_i)$, displayed as a block in the center of each diagram, which can always be synthesized by two CNOT gates and two one-qubit gates, see figure \ref{fig:block_form}. For each $\otimes\sigma_x$ two CNOT gates are added, where the control qubit is always the $n$th dimension and the target qubit is given by the $i$th dimension of the subalgebra $\mathfrak{h}(i)$ it gets generated from. The control qubits of the rest of the CNOT gates enlarge up to the respective $n$th dimension, with the exception of the outermost CNOT gates, which remain unaltered since they serve as a final permutation.}
	\label{fig:h4_recursive}
\end{figure*}

\begin{figure}[h]
\includegraphics{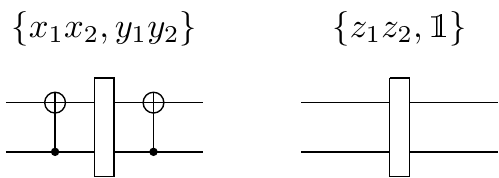}

\caption{Quantum circuits generating the exponentials of the generators of the Lie subalgebra $\mathfrak{h}(2)$ by using a block-diagonal form. Each of the central blocks contains an instance of the dimensionally adapted circuit shown in figure \ref{fig:block_form}.}
\label{fig:a(2)}
\end{figure}

The method we employ to decompose the exponential terms $M_i(a_i,b_i)$ and $N_i(a_i,b_i)$ is through a block-diagonal matrix $P_\mp(a,b)$. The method we found, based on the recursive algorithm (\ref{eq:recursion}), starts by grouping the generator $\mathfrak{a}(2)$ into its proportional terms
\begin{equation}
    \mathfrak{a}(2)=i\{x_1x_2,y_1y_2\}\bigcup i\{z_1z_2,\mathbbm{1}\}
\end{equation}
The circuits corresponding to these algebra elements are shown in figure \ref{fig:a(2)} which decomposes the block-diagonal matrix $P_\mp$. Expansion to larger elements and therefore higher dimensional structures can be done recursively through adding more terms in the algebra. The central $P_\mp$ element expands in dimension alongside.

\begin{figure}[h]
	\includegraphics{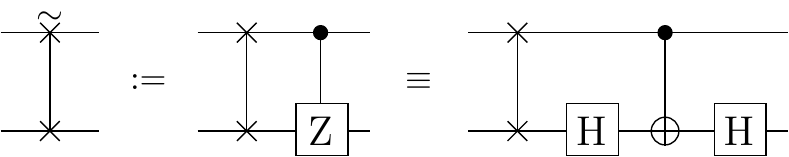}
	\caption{Quantum circuit decomposing a fermionic SWAP gate. A fermionic SWAP can be decomposed at worst through four CNOT gates.}
\label{fig:fSWAP}
\end{figure}

\begin{figure}[h]
	\includegraphics{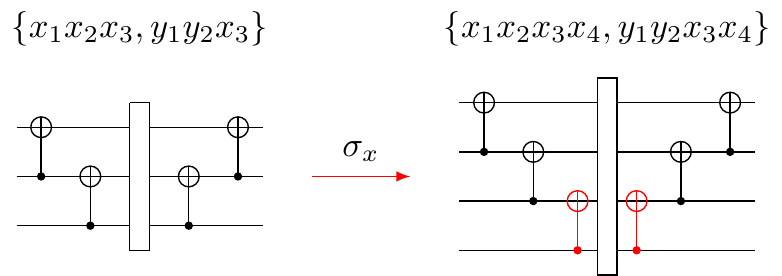}
	\includegraphics{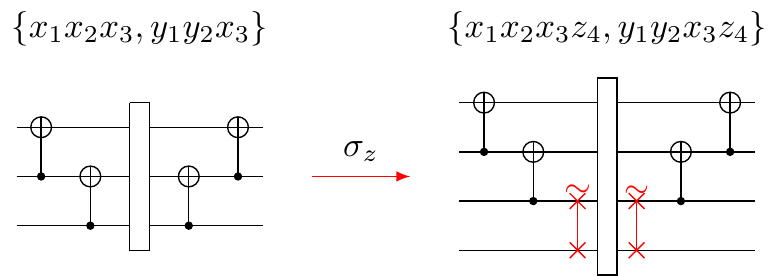}
	\caption{Quantum circuits decomposing the exponential terms $M_1(a_1,b_1)=e^{i(a_1x_1x_2x_3x_4+b_1y_1y_2x_3x_4)}$ (top) and $N_2(a_2,b_2)=e^{i(a_2x_1x_2x_3z_4+b_2y_1y_2x_3z_4)}$ (bottom). The action of $\otimes \sigma_x$ introduces two additional CNOT gates and the action of $\otimes \sigma_z$ adds two additional SWAP gates, whose target qubits are given by the subalgebra $\mathfrak{h}(3)$. The rest of the control qubits in $\mathfrak{h}(4)$ enlarge up to the $4$th dimension with the exception of the outermost CNOT gates, which serve only as a final diagonal permutation.}
\label{fig:explanation_x1x2x3x4}
\end{figure}

\begin{figure*}[hbt]
	\includegraphics{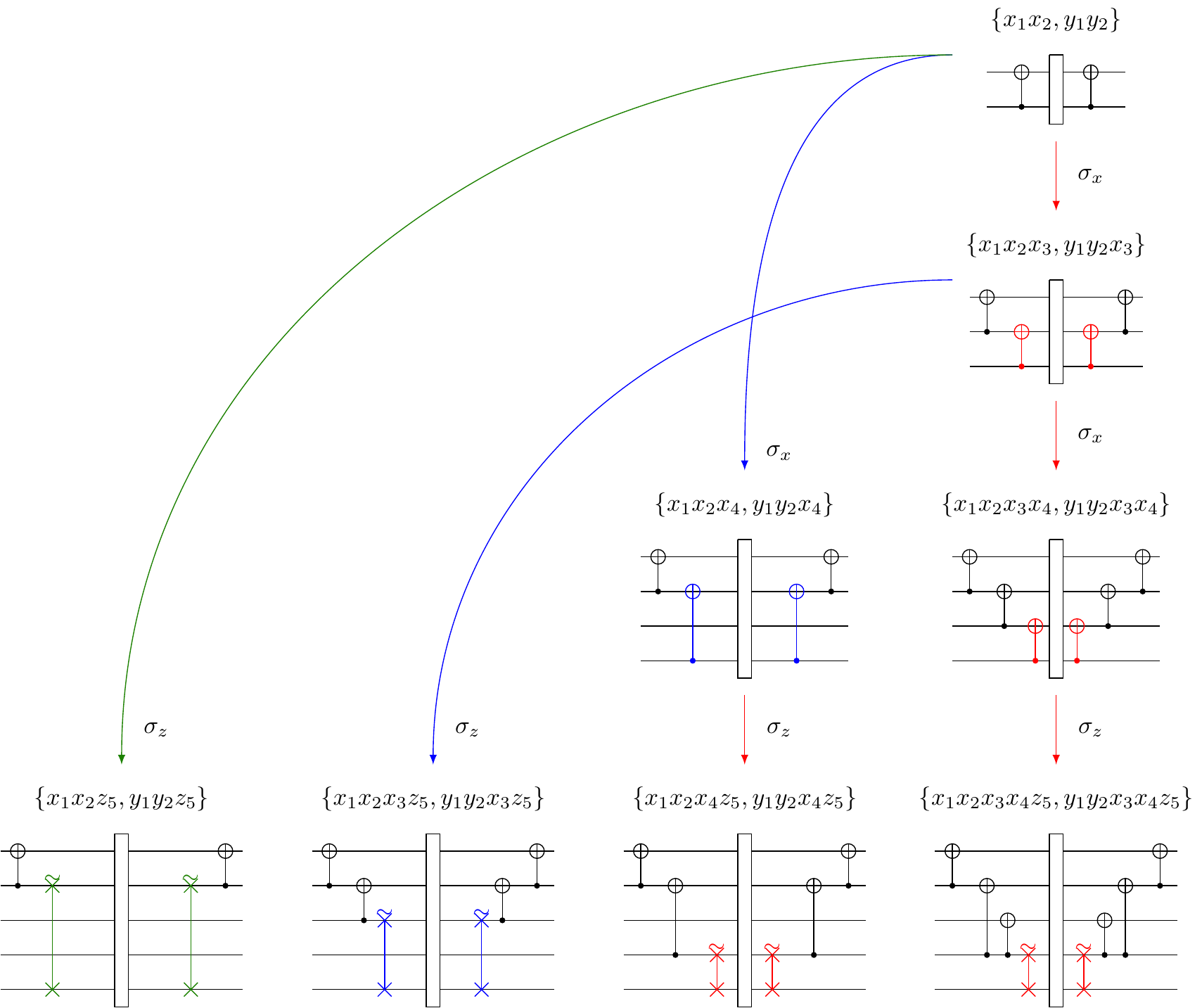}
	\caption{Recursive algorithm displaying how a part of the Cartan subalgebra $\mathfrak{f}(5)$ gets generated. For each $\otimes\sigma_x$ two CNOT gates are added, while for each $\otimes\sigma_z$ two SWAP gates are added. The control qubit of the additional gates is always the $n$th dimension, while the target qubit is given by the $i$th dimension of the subalgebra $\mathfrak{h}(i)$ it is generated from, with the exception of the outermost always unaltered CNOT gates, which serve only as a final diagonal permutation. Moreover, for each $\otimes\sigma_x$ the control qubit of the rest of the CNOT gates enlarges up to the respective $n$th dimension.}
	\label{fig:recursive_h}
\end{figure*}

We can differentiate expansion into higher dimensions along $\sigma_x$ and $\sigma_z$. We find that there is a direct correspondence between enlarging the algebra and the circuit construction. Adding a $\sigma_x$ to the algebra corresponds to adding a CNOT gate from the $n$th to the $n-1$th quantum gate. For $\sigma_z$, the corresponding gate is a fermionic SWAP gate, see figure \ref{fig:fSWAP}, between the same wires. This gives a circuit construction as shown in figure \ref{fig:explanation_x1x2x3x4} for $SU(16)$. Higher dimensions work in a similar fashion and exhibit a branching structure depending on which algebra dimensions are added. This is shown in more detail in figures \ref{fig:h4_recursive} and \ref{fig:recursive_h}. The structure is also relevant for the CNOT count, where each fermionic SWAP will count in the end as one CNOT gate, see figure \ref{fig:fSWAP}.

%    \begin{figure}[h]
%
%	\caption{Quantum circuits converting the exponential term $M(a,b)=e^{i(ax1x2x3z4+by1y2x3z4)}$ into a block-diagonal form. The action of $\otimes \sigma_z$ adds two additional SWAP gates whose target qubits are given by the subalgebra $\mathfrak{h}(3)$.}
%	\label{fig:explanation_x1x2x3z4}
%    \end{figure}\noindent

\begin{figure}[h]
	\centering
	\includegraphics{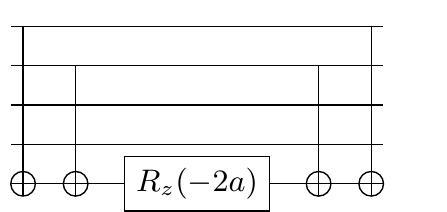}	\caption{Quantum circuit decomposing the exponential term $N(a)=e^{iaz_1z_2z_5}\equiv \text{Diag}(z_1z_2z_5)\in SU(32)$. This quantum circuit appears for all $n\geq 3$ dimensions, where the target qubits and the one-qubit gate have to be adjusted accordingly to the dimension of the quantum circuit.}
	\label{fig:diagonal_form}
\end{figure}

In addition to these structures, on every dimension $n\geq 3$, there is one generator $z_1z_2z_n$ of the Cartan subalgebra $\mathfrak{f}(n)$ which is more efficient to treat separately. We found  that such an exponential term depending on one parameter can always be decomposed, regardless of the dimension, with a single rotation gate surrounded by four dimensionally adapted CNOT gates, see figure \ref{fig:diagonal_form}.

These constructions are sufficient to implement all possible algebra generators since they cover the whole subalgebra given in definition \ref{def:cartan-subalgebra}. Hence all possible unitaries $U\in SU(2^n)$ can be covered by the construction.

\section{Example}\label{sec:example}

We now apply the decomposition method described above to an operator $U\in SU(8)$. Using the Cartan decomposition (\ref{eq:cartandecomposition_corollary2}) the following decomposition is obtained
\begin{equation*}
U=K_1\exp(z_1)K_2\exp(y)K_3\exp(z_3)K_4,
\end{equation*}
where $K_i\in SU(4)\otimes U(1)$, and $z\in\mathfrak{h}(3)$ and $y\in\mathfrak{f}(3)$, where $\mathfrak{h}(3)$ is a Cartan subalgebra of $(\mathfrak{su}(8),\mathfrak{su}_{\mathfrak{l}}(8))$ and $\mathfrak{f}(3)$ is a Cartan subalgebra of $(\mathfrak{su}_{\mathfrak{l}}(8),\mathfrak{su}_{\mathfrak{l0}}(8))$.
The elements of the Lie subalgebra are generated by (\ref{eq:recursion}) and thus given by
\begin{equation*}
    \mathfrak{h}(3)=\text{span}\hspace{1ex}i\{x_1x_2x_3, y_1y_2x_3\}\bigcup i\{z_1z_2x_3,x_3\},
\end{equation*}
\begin{equation*}
    \mathfrak{f}(3)=\text{span}\hspace{1ex}i\{x_1x_2z_3, y_1y_2z_3\}\bigcup i\{z_1z_2z_3\}
\end{equation*}
where we have already grouped the proportional terms.

By means of the recursive algorithm introduced previously, we decompose the exponential terms of the generators of this subalgebra through a quantum circuit including a block-diagonal form $P_\mp(a,b)$. For instance, the unitary $M_1(a_1,b_1)$ defined by
\begin{equation}
    M_1(a_1,b_1)=e^{i(a_1x_1x_2x_3+b_1y_1y_2x_3)}
\end{equation}
which denotes the exponential of the generators $\{x_1x_2x_3, y_1y_2x_3\}$, is decomposed through the quantum circuit shown in figure \ref{fig:recursive_h3} below, where the white box denotes the block-diagonal operator $P_\mp(a_1,b_1)$. 

\begin{figure}[h]
	\includegraphics{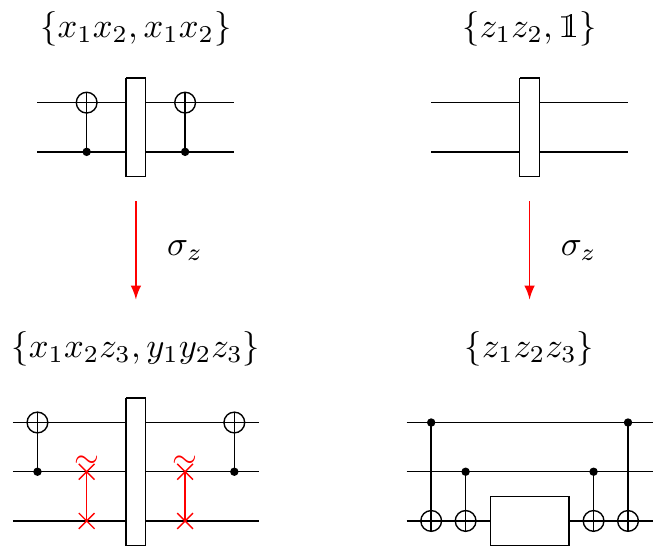}
	\caption{Recursive algorithm showing how the exponential elements coming from the terms in the subalgebra $\mathfrak{h}(2)$ by tensoring with $\sigma_z$ are generated by adding a fermionic SWAP gate on each side. However, the exponential term which comes from $\{z_1z_2z_3\}$ and depends only on one parameter has to be treated separately, see figure \ref{fig:diagonal_form}.}
	\label{fig:recursive_f3}
\end{figure}

The rest of the generators can be decomposed by following the same algorithm introduced in the previous section, see figures \ref{fig:recursive_h3} and \ref{fig:recursive_f3}. The exponential term involving the generator $\{z_1z_2z_3\}$ can always be generated by an analogous quantum circuit as the one shown in figure \ref{fig:diagonal_form} involving four CNOT gates and one one-qubit gate. The entire construction can be seen in figure \ref{fig:decomposition-representation}.

\begin{figure}[h]
	\includegraphics{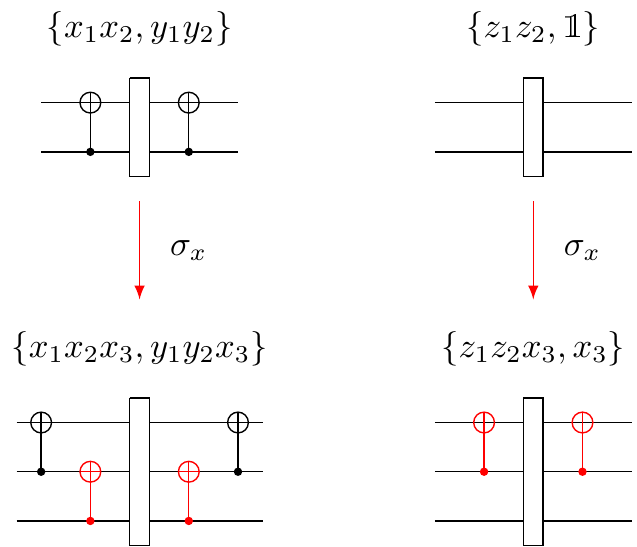}	
	\caption{Recursive algorithm showing how to decompose the unitaries $M_1(a_1,b_1)=e^{i(a_1x_1x_2x_3+b_1y_1y_2x_3})$ and $M_2(a_2,b_2)=e^{i(a_2z_1z_2x_3+b_2x_3})$  generated from the respective subalgebra elements $\{x_1x_2,y_1y_2\}$ and $\{z_1z_2,\mathbbm{1}\}$ of $\mathfrak{a}(2)$ by tensoring with $\sigma_x$. For every tensor product with a $\sigma_x$ matrix, there is an additional CNOT gate on each side whose target qubit is given by $\mathfrak{h}(2)$, with the exception of the outer CNOT gates since they serve only as a final diagonal permutation.}
	\label{fig:recursive_h3}
\end{figure}

By just counting we can see that there are six CNOT gates and two block-diagonal $P_\mp$ matrices decomposing the exponential terms $M_1(a_1,b_1)=e^{i(a_1x_1x_2x_3+b_1y_1y_2x_3})$ and $M_2(a_2,b_2)=e^{i(a_2z_1z_2x_3+b_2x_3})$. To decompose the exponential terms $N_1(c_1,d_1)=e^{i(c_1x_1x_2z_3+d_1y_1y_2z_3)}$ and $N_2(c_2)=e^{ic_2z_1z_2z_3}$ there are six CNOT gates, two fermionic SWAP gates, and one block-diagonal $P_\mp$ matrix. Therefore, to decompose a unitary U in $SU(8)$ we need a total of 54 CNOT gates, where we have assumed that every two-qubit circuit can be decomposed at most by three CNOT gates and every fermionic SWAP gate by at most three CNOT gates. It is possible to get rid of the SWAP gates by interchanging the roles of the second and the third qubit and thus reduce the number of CNOT gates to 42. 

Although this is slightly worse than the previous unstructured Cartan decomposition method \cite{vatan2004realization} of a three-qubit unitary which required a total of 40 CNOT gates, it is possible to further improve the cost to 38 CNOT gates by absorbing some of the CNOT gates by the neighboring $A_i\in SU(4)$ unitaries, which can be seen representated in figure \ref{fig:decomposition-representation}.

\section{Number of CNOT gates}\label{sec:cnot_gates}

In order to determine the amount of CNOT gates required, let $\mathcal{C}_n$ denote the number of CNOT gates coming directly from the diagram decomposition for $SU(2^n)$, where $\mathcal{C}_{\mathfrak{h}(n)}$ specifies the number of CNOT gates required to synthesize the exponential operator $M=e^y$, and $\mathcal{C}_{\mathfrak{f}(n)}$ specifies the number of CNOT gates required to synthesize the exponential operator $N_i=e^{z_i}$. From \ref{cor:4group} follows then $\mathcal{C}_{n}=\mathcal{C}_{\mathfrak{h}(n)}+2\mathcal{C}_{\mathfrak{f}(n)}$. Since the amount of CNOT gates in the recursive algorithm of the Cartan subalgebra $\mathfrak{h}(n)$ follows the same structure as Pascal's triangle, $\mathcal{C}_{\mathfrak{h}(n)}$ is given by the following equation
\begin{equation}
	\mathcal{C}_{\mathfrak{h}(n)}=\sum_{k=0}^{n-2}\binom{n-2}{k}2(k+1)\text{CNOT}+2^{n-2}P_\mp
\end{equation}
where $P_\mp$ denotes the block-diagonal matrices, which always consist of two CNOT gates and two one-qubit gates, see figure \ref{fig:block_form}. 
%\begin{remark}
%    This can also be justified since $\mathfrak{f}$ is a Cartan subalgebra of $(\mathfrak{su}_l(2^n),\mathfrak{su}_{l0}(2^n))$, which gets decomposed by the sum of $\mathfrak{su}_{l0}$ and $\mathfrak{su}_{l1}$, where the former gets spanned by tensoring with $\sigma_z$ and the latter by tensoring with the identity. By manually getting rid of the SWAP gates amounts to expand the Lie algebra $\mathfrak{b}(n)$ in (\ref{eq:recursion}) by tensoring with $\mathbbm{1}$ rather than $\sigma_z$. However, we lose the maximally Abelian condition, and thus $\mathfrak{f}(n)$ is not a Cartan subalgebra anymore.
%\end{remark}
To count the number of CNOT gates for $\mathcal{C}_{\mathfrak{f}(n)}$ note that the fSWAP gate consists of a SWAP gate followed by two Hadamard gates and one CNOT. Note that all occurring fSWAPS are adjacent to the block-diagonal matrix $P_\mp$ and therefore we can get rid of the internal SWAPS by manually adjusting the one-qubit and the dimension of the target qubits of the block-diagonal matrix. Thus, in terms of CNOTs, adding a pair of fSWAPS effectively introduces two CNOTS. Moreover, the number of CNOT gates in the recursive algorithm of $\mathfrak{f}(n)$ also follows the structure of Pascal's triangle,
\begin{equation}
\begin{split}
    		\mathcal{C}_{\mathfrak{f}(n)}=&\sum_{k=1}^{n-2}\binom{n-2}{k}2(k+1)\text{CNOT}\\
      &+(2^{n-2}-1)P_\mp+\text{Diag}(z_1z_2z_n)
\end{split}
\end{equation}
where $\text{Diag}(z_1z_2z_n)$ denotes the generator $z_1z_2z_n$ that is not proportional to any other generator and which always consists, regardless of the dimension, of four CNOT gates and one-qubit gate, see figure \ref{fig:diagonal_form}.

Hence, the number of CNOT gates $\mathcal{C}_{n}$ required to synthesize the exponential operators is
\begin{equation}
    \begin{split}
        \mathcal{C}_{n}=&6\sum_{k=0}^{n-2}\binom{n-2}{k}(k+1)\text{CNOT}\\
        &+3\left(2^{n-2}-\frac{2}{3}\right)P_\mp+2\text{Diag}(z_1z_2z_n)-4\\
        =&6\sum_{k=0}^{n-2}\binom{n-2}{k}\left(k+1\right)+ \frac{3}{2} 2^n
    \end{split}
\end{equation}
where we used the fact that every block-diagonal matrix $P_\mp$ can be decomposed by two CNOT gates, see figure \ref{fig:block_form}. This binomial sum can be determined by means of the following identities
\begin{equation}
	\sum_{k=0}^{n}\binom{n}{k}=2^n,\hspace{3ex} \sum_{k=0}^{n}k\binom{n}{k}=n2^{n-1}
\end{equation}
Therefore, $\mathcal{C}_{n}$ is given by
\begin{equation}
    \begin{split}
        \mathcal{C}_{n}&= 3 \cdot \left( 2^{n-1}+n2^{n-2} \right)
        \label{eq:cnot}
    \end{split}
\end{equation}
To determine the entire number of CNOT gates for a unitary $U\in SU(2^n)$ we also need to take into consideration the CNOT gates that recursively come from lower dimensions, see equation (\ref{eq:cartandecomposition_corollary2}) and its corresponding figure \ref{fig:decomposition-representation}. To that end, let $T_n$ denote the total number of CNOT gates for a unitary $U\in SU(2^n)$. By (\ref{cor:4group}), which follows from the decomposition of a unitary, we have that the total number of CNOT gates for a unitary $U$ in $SU(2^n)$ is given by
\begin{equation}
\begin{split}
    T_n&=\mathcal{C}_n+4\mathcal{C}_{n-1}+4^2\mathcal{C}_{n-2}+...=\sum_{i=2}^{n}4^{n-i}\mathcal{C}_i\\
    &=4^{n-2}\mathcal{C}_2+ 3 \cdot 2^{n-3}\left[3 \cdot 2^n-2n-8\right]
\end{split}
 \end{equation}
which was determined by using equation (\ref{eq:cnot}) and where we have explicitly separated $\mathcal{C}_2$ since our $n=2$ base case does not work recursively. In \cite{vatan2004optimal}, \cite{vidal2004universal}, and \cite{shende2004minimal} it was proven that a two-qubit quantum circuit could usually be synthesized with at most three CNOT gates. Therefore, the total number of CNOT gates required to decompose a unitary $U$ in $SU(2^n)$ by means of the Khaneja-Glaser decomposition algorithm is
\begin{equation}
\begin{split}
    T_n&=\frac{3}{16}4^n+ \frac{9}{8} 4^n -3\left(n2^{n-2}+2^{n}\right)\\
    &= \frac{21}{16}4^n-\mathcal{O}(n2^n)
\end{split}
\end{equation}
which is roughly a factor of five away from the best-known theoretical lower bound for synthesizing an $n$-qubit unitary, \cite{shende2004minimal}. As our method is recursive and creates unitaries of any $(n-k)$ qubits size, more optimal unitary decompositions for a particular number of qubits can be taken into account and inserted at that size.

\section{Discussion}\label{sec:Discussion}

We have implemented the algorithm in rudimentary form and provide it in the supplementary material \cite{supplementary}. The algorithm can certainly be optimized further. We leave this as implementation work for colleagues more familiar with suitable programming environments. 

The presented decomposition algorithm provides a solid basis for decomposing any arbitrary unitary in $\operatorname{SU}(2^n)$. For the construction, we assume a ideal quantum computer with any-to-any connections and no noise. This assumption is common to circuit constructions algorithms and can be remedied by post-creation optimization of the circuit. The first assumption can be approached either by exchanging CNOTs on non-existing connections with CNOT ladder chains that implement an equivalent operation. In the worst case of a linear chain, a CNOT connecting qubits $k$ apart, $4(k-1)$ nearest neighbour CNOTs are required \cite{shende2005synthesis}. It may be possible to optimize this through our approach, since there is a direct correspondence between the subalgebra generators and the CNOT gates between qubits. Restricting the subalgebra to exclude certain connections may provide a more optimal solution. We suggest this approach for future work.

The robustness of a circuit with respect to noise is much more difficult to measure and achieve. The current methods for achieving fault-tolerance, such as stabilizer codes \cite{webster_universal_2022} and logical qubits \cite{marques_logical-qubit_2022} are not easily implementable in our approach. However, it is possible to create an near-optimal circuit with the presented method and adjust it to be noise-tolerant afterwards.

%As mentioned in the three-qubit unitary decomposition, it is possible to optimize the number of CNOT gates in each dimension $n$ by absorbing the CNOT gates that are adjacent to the $(n-1)$-qubit unitaries $A_i$, see figure \ref{fig:decomposition-representation}. Thus, a future optimization process might be to implement an algorithm capable of counting how many CNOT gates might be absorbed in each dimension. Moreover, there might also be more efficient circuits synthesizing a particular exponential operator, although these circuits might arise as a special case and do not generalize recursively. 

Our approach is not limited to CNOT gates. While we use it throughout our construction, it can be readily transformed to another gateset, as long as it forms a universal family of quantum gates \cite{barenco1995elementary}. Thus, if the particular hardware can only (or efficiently) implement a different set of control gates, it is possible to translate the circuit into a different family of universal gates. That is, the methods described here generalize to different families of universal quantum gates, which might be more easily implemented on the particular quantum hardware.

What sets apart the Khaneja-Glaser Cartan decomposition of a unitary described here from the other decomposition methods is that it gives an explicit construction of the quantum circuit decomposing a unitary and thus can be directly implemented on a quantum computer. Moreover, this decomposition method can also be used to optimize existing computational circuits to improve their scaling.

\begin{table}
    \centering
    \def\arraystretch{1.5}
    \setlength\tabcolsep{6pt}
    \begin{tabular}{ll}

    Algorithm& CNOT gate count\\
    \hline
         Original decomp. \cite{barenco1995elementary}& $\mathcal{O}(n^34^n)$ \\
         Asymptotic decomp. \cite{knill1995approximation}&$\mathcal{O}(n4^n)$\\ 
         Gray codes \cite{vartiainen2004efficient}&$\mathcal{O}(4^n)$\\
         Cosine-Sine decomp. \cite{mottonen2004quantum}&$4^n-2^{n+1}$\\
         Optimized QSD \cite{shende2005synthesis}& $\frac{23}{48}4^n-\frac{3}{2}2^n+\frac{4}{3}$\\  
         KG Cartan decomposition&$\frac{21}{16}4^n-3\left(n2^{n-2}+2^{n}\right)$\\ \hline
         Theoretical lower bound \cite{shende2004minimal}&$\lceil\frac{1}{4}\left(4^n-3n-1\right)\rceil$
    \end{tabular}
    \caption{Comparison of different methods for the number of CNOT gates necessary for synthesizing an $n$-qubit unitary. The Khaneja-Glaser decomposition of a unitary is a factor of four from the lower theoretical bound.}
    \label{tab:comparison_CNOT}
\end{table}

The method presented in this paper demonstrates how to efficiently build quantum circuits implementing an $n$-qubit unitary operation through the Cartan decomposition of Lie algebras. Our work generalizes the previous unstructured Cartan decomposition of a three-qubit unitary to a structured recursive algorithm capable of synthesizing any desired unitary operation. Our construction allows the expansion of any quantum circuit  in terms of rotation matrices and generators.  Moreover, we show how these generators can be recursively decomposed through CNOT and fermionic SWAP gates into circuits that can be directly implemented on a quantum computer. This Cartan decomposition method also scales well, with a near-optimal scaling of $\frac{21}{16}4^n-3\left(n2^{n-2}+2^{n}\right)$ CNOT gates required to synthesize an $n$ qubit unitary operation.
The algorithmic structure of the method and constructions described in this paper allows for a simple yet flexible implementation, both in terms of applications of the algorithm and software and hardware architectures. %Thus, this method provides a useful bridge between quantum algorithms and quantum hardware.

\begin{acknowledgements}

The authors acknowledge funding from the German Federal Ministry of Education and Research (BMBF) under the funding program ``Förderprogramm Quantentechnologien – von den Grundlagen zum Markt" (funding program quantum technologies – from basic research to market), project BAIQO, 13N16089.

\end{acknowledgements}

\end{document}